\documentstyle[aps,prb,twocolumn]{revtex}

\input epsf

\begin{document}

\draft

% THE FOLLOWING GIVES A FULL WIDTH TITLE AND ABSTRACT.  IT IS COUPLED
% WITH A LINE BELOW.  COMMENT-OUT BEFORE SUBMISSION.
\twocolumn[\hsize\textwidth\columnwidth\hsize\csname@twocolumnfalse%
\endcsname

\title{Single-hole properties in the $t$-$J$ and strong-coupling
models}

\author{B. M. Elrick\cite{email} and A. E. Jacobs\cite{email}}
\address{Department of Physics, University of Toronto,
         Toronto, Ontario, Canada M5S 1A7}
\date{\today}
\maketitle

\begin{abstract}
We report numerical results for the single-hole properties in the
$t$-$J$ model and the strong-coupling approximation to the Hubbard
model in two dimensions.  Using the hopping basis with over $10^6$
states we discuss (for an infinite system) the bandwidth, the leading
Fourier coefficients in the dispersion, the band masses, and the
spin-spin correlations near the hole.  We compare our results with
those obtained by other methods.  The band minimum is found to be at
($\pi/2,\pi/2$) for the $t$-$J$ model for $0.1 \leq t/J \leq 10$, and
for the strong-coupling model for $1 \leq t/J \leq 10$.
The bandwidth in both models is approximately $2J$ at large $t/J$, in
rough agreement with loop-expansion results but in disagreement
with other results.
The strong-coupling bandwidth for $t/J\agt6$ can be obtained from the
$t$-$J$ model by treating the three-site terms in
first-order perturbation theory.
The dispersion along the magnetic zone face is flat, giving a large
parallel/perpendicular band mass ratio.
\end{abstract}

%71.27.+a Strongly correlated electron systems
%71.28.+d Narrow-band systems; heavy-fermion solids;
%         intermediate-valence solids
%74.25.Jb Electronic structure
%74.72.-h High-T sub c cuprates
%75.10.Jm Quantized spin models

% THE FOLLOWING GIVES A FULL WIDTH TITLE AND ABSTRACT.  IT IS COUPLED
% WITH A LINE ABOVE.  COMMENT-OUT BEFORE SUBMITTING.
]

\narrowtext

\section{Introduction}
Anderson's suggestion\cite{Anderson_1987_1196} that the copper-oxygen
planes of the high-temperature superconductors\cite{Bednorz_1986_189}
are strongly correlated systems has sparked renewed interest in the
two-dimensional Hubbard model.  Much of our understanding of the
strong-coupling limit of the model, and the related $t$-$J$ model, has
been obtained by numerical work (reviewed in
Ref.~\onlinecite{Dagotto_1994_763}).  Although the single-hole
properties have been studied extensively, exact-diagonalization
studies of small systems are hindered by large finite-size effects in
the parameter region of interest, and Monte-Carlo studies of larger
systems are hindered by the minus-sign problem; other methods have
also been used, but there is still no general agreement on these
properties, particularly for $t/J$ values in the physical region.  For
this reason, we have studied the single-hole properties using the
hopping basis of Trugman\cite{Trugman_1988_1597,Trugman_1990_892} and
compared them with results obtained by other methods.

In the limit $U\gg t$, the Hubbard Hamiltonian can be approximated by
the strong-coupling Hamiltonian\cite{Hirsch_1985_1317} $H_{sc} =
H_{t\mbox{\small -}J} + H_3$; this differs from the $t$-$J$
Hamiltonian $H_{t\mbox{\small -}J}$ (which has its own justification)
by the three-site terms in $H_3$:
\begin{eqnarray}
\label{t-J_Hamiltonian_eq}
H_{t\mbox{\small -}J} &=&
  -t \sum_{i,\delta,\sigma}
    c_{i+\delta,\sigma}^\dagger c_{i,\sigma}^{\phantom\dagger}
  + J \sum_{\langle ij\rangle}({\bf S}_i \cdot {\bf S}_j
                               - \textstyle {1\over 4} n_i n_j)\ , \\
\label{three_site_term_eq}
H_3  &=&  -{{J}\over{4}}\sum_{i,\sigma}\sum_{\delta,\delta^\prime}
(c_{i+\delta,\sigma}^\dagger  c_{i,-\sigma}^\dagger
c_{i,-\sigma}^{\phantom\dagger}c_{i+\delta^\prime,\sigma}^{\phantom\dagger}
\nonumber \\
& & \qquad\qquad\ \ -c_{i+\delta,-\sigma}^\dagger c_{i,\sigma}^\dagger
c_{i,-\sigma}^{\phantom\dagger}c_{i+\delta^\prime,\sigma}^{\phantom\dagger})\ ;
\end{eqnarray}
here sites $i+\delta$ and $i+\delta^\prime$ are distinct nearest
neighbors of site $i$, $\langle ij \rangle$ are nearest-neighbor
pairs, and $J = 4t^2/U$.  The $t$-$J$ and strong-coupling Hamiltonians
operate in the reduced Hilbert space with no doubly occupied sites;
this restriction is implicit in the above.  Validity of the
strong-coupling approximation requires $U\gg t$; the parameter range
believed appropriate to the high-temperature superconductors is
$2<t/J<10$, or $8<U/t<40$.  We present results for the $t$-$J$ model in
the region $0.1 \leq t/J \leq 10$ and for the strong-coupling model in
the region $1 \leq t/J \leq 10$.

The single-hole properties in the $t$-$J$ and strong-coupling models
have been studied previously, the first having received more
attention.  Methods include exact-diagonalization studies of small
lattices\cite{Kaxiras_1988_656,Dagotto_1989_6721,Dagotto_1990_9049,%
Bonca_1989_7074,Elser_1990_6715,Roder_1991_6284,Fehske_1991_8473,%
Elrick_1993_6004}, studies of infinite lattices using a restricted
basis set\cite{Trugman_1988_1597,Trugman_1990_892,Inoue_1990_2110,%
Inoue_1990_3467,Inoue_1994_6213,Macready_1991_5166}, Monte-Carlo
methods\cite{Boninsegni_1991_10353,Boninsegni_1992_560,Boninsegni_1992_4877,%
Dagotto_1991_8705,Giamarchi_1993_2775,Barnes_1993_11247,Dagotto_1994_728},
and other
methods\cite{Schmitt-Rink_1988_2793,Kane_1989_6880,Marsiglio_1991_10882,%
Martinez_1991_317,Liu_1991_2414,Liu_1992_2425,Liu_1995_3156,%
Sushkov_1992_303,Sushkov_1992_199}.  Properties discussed include the
ground-state energy, the bandwidth, the dispersion, the band masses,
the nearest-neighbor spin-spin correlations and the spectral function.
% KEEP THE FOLLOWING AS A RESOURCE
%energy\cite{Dagotto_1989_6721,Dagotto_1990_9049,Bonca_1989_7074,%
%Elser_1990_6715,Roder_1991_6284,Fehske_1991_8473,Elrick_1993_6004,%
%Inoue_1990_2110,Macready_1991_5166,Boninsegni_1991_10353,%
%Boninsegni_1992_560,Boninsegni_1992_4877,Giamarchi_1993_2775,%
%Barnes_1993_11247,Marsiglio_1991_10882,Liu_1991_2414,Liu_1992_2425,%
%Liu_1995_3156,Sushkov_1992_303}, the
%dispersion\cite{Dagotto_1990_9049,Elser_1990_6715,Roder_1991_6284,%
%Fehske_1991_8473,Inoue_1990_2110,Inoue_1994_6213,Macready_1991_5166,%
%Boninsegni_1991_10353,Giamarchi_1993_2775,Barnes_1993_11247,%
%Marsiglio_1991_10882,Liu_1992_2425,Liu_1995_3156,Sushkov_1992_199},
%the bandwidth\cite{Dagotto_1990_9049,Bonca_1989_7074,%
%Marsiglio_1991_10882,Liu_1992_2425,Liu_1995_3156}, the band
%masses\cite{Elser_1990_6715,Fehske_1991_8473,Kane_1989_6880,%
%Marsiglio_1991_10882}, the spin-spin correlation
%functions\cite{Dagotto_1989_6721,Elser_1990_6715,Inoue_1990_3467}, and
%the spectral function\cite{Dagotto_1990_9049,Macready_1991_5166,%
%Schmitt-Rink_1988_2793,Kane_1989_6880,Liu_1991_2414,Liu_1992_2425,%
%Liu_1995_3156,Marsiglio_1991_10882}.
As well, there is an extensive literature on the Hubbard model itself,
including recent finite-temperature Monte-Carlo
results\cite{Bulut_1994_705,Bulut_1994_748,Bulut_1994_7215}.

This paper studies the one-hole properties on an infinite lattice,
using a restricted basis set (in effect a variational method).
Section \ref{hopping_basis} describes the basis, and Sections
\ref{dispersion}-\ref{correlations} give results for the bandwidth,
the dispersion, the band masses, and the nearest-neighbor spin-spin
correlations respectively.  For both models, the band minimum is at
${\bf k}=(\pi/2,\pi/2)$ and the maximum at ${\bf k}=(0,0)$ for the
$t/J$ values investigated.  The bandwidth is approximately $2J$ at
large $t/J$, in agreement with loop-expansion
results\cite{Marsiglio_1991_10882,Martinez_1991_317,Liu_1995_3156} and
in disagreement with variational Monte-Carlo
results\cite{Boninsegni_1992_4877}.  At large $t/J$, the effects of
three-site terms on the bandwidth are well described by first-order
perturbation theory using the $t$-$J$ ground-state wavefunction; that
is, the three-site terms appear to have little effect on the
ground-state wavefunction at large $t/J$.  The band mass parallel to
the zone face is much larger than the perpendicular mass.  The
spin-spin correlations near the hole are reduced relative to the
starting state, but remain antiferromagnetic.

\section{Hopping Basis}
\label{hopping_basis}

We study a system of $N-1$ electrons on a square lattice of $N$ sites
with periodic boundary conditions; the Hilbert space is restricted to
the $S_z=1/2$ sector with no doubly occupied sites.  We use the same
basis for both models, namely the hopping
basis\cite{Trugman_1988_1597,Trugman_1990_892} which has been used
previously\cite{Trugman_1988_1597,Trugman_1990_892,Inoue_1990_2110,%
Inoue_1990_3467,Inoue_1994_6213,Macready_1991_5166}.  This method
allows the study of infinite systems (eliminating finite-size
effects), but only certain properties, like the bandwidth and the band
masses, can be studied.

In zeroth order, the basis (denoted $B_0$) consists of a single state
(denoted $|cN\rangle$), the N\'eel state with a missing down-spin
electron.  Higher-order bases are generated by repeatedly applying the
$t$ term in the Hamiltonian (which hops the hole to a nearest-neighbor
site).  The first-order basis $B_1$ contains the $|cN\rangle$ state
plus the four states generated by hopping the hole.  The $n$-th order
basis $B_n$ consists of the states in the basis $B_{n-1}$ plus those
generated by applying the hopping operator to the states in the
difference $B_{n-1}-B_{n-2}$.  The basis size (values are given in
Table \ref{basissize_table}) grows exponentially with order.  The
hopping basis, which emphasizes states differing from the $|cN\rangle$
state only near the hole, cannot give a good value of the ground-state
energy (because, for example, it does not generate spin interchanges
far from the hole in reasonable order); the expectation is that it
describes well properties like the dispersion and the nearest-neighbor
spin-spin correlations near the hole.

We have used the bases from $B_6$ to $B_{13}$ for most quantities,
going to such large bases because some properties were still changing
significantly; even with basis $B_{13}$ ($\sim2\times10^6$ states),
however, some properties are incompletely converged.  Various
extrapolation schemes were considered but judged unreliable, and so we
usually present values for the three largest bases to provide an
estimate of the error due to the truncation of the basis.
% Previous work\cite{Inoue_1990_2110,Inoue_1990_3467} used the bases up
% to $B_5$.  There is a discrepancy in the number of states in $B_5$
% between this work and ours.

The system size ($16\times16$; the lattice constant $a$ is unity) is
effectively infinite since there are no paths which wrap around the
system even in 13-th order.  Since the hole moves in an
antiferromagnetic background, the Brillouin zone is reduced to the
square formed by the points $(\pm\pi,0)$ and $(0,\pm\pi)$.  The
symmetries of the lattice reduce the independent part of the Brillouin
zone to the triangle with corners at $(0,0)$, $(\pi,0)$, and
$(\pi/2,\pi/2)$, denoted $\bf\Gamma$, $\bf M$, and $\bf S$
respectively.  Each state $|n\rangle$ in the basis is a Bloch state,
an eigenstate of the translation operator corresponding to an allowed
value of the momentum.  For each basis, and each value of the momentum
{\bf k}, the lowest eigenvalue and eigenvector were found using a
conjugate-gradient method to minimize the function
$\langle\Psi|H|\Psi\rangle/\langle\Psi|\Psi\rangle$ with respect to
the expansion coefficients in $|\Psi\rangle=a_n|n\rangle$; this method
is reported to converge more rapidly than others commonly
used\cite{Nightingale_1993_7696}, but gives the eigenvector to only
single precision. Where necessary, the eigenvector was improved by a
Lanczos method.

The dispersion (in the energy as a function of {\bf k}) results from
several processes.  The Trugman
paths\cite{Trugman_1988_1597,Trugman_1990_892} translate the hole to a
next-nearest-neighbor site or a third-nearest-neighbor site on the
same sub-lattice, restoring the original configuration. In the
lowest-order path, the hole hops six times around the smallest square
to a next-nearest-neighbor site; as a result, matrix elements like
$\langle B_2 | c_{i+\delta,\sigma}^\dagger
c_{i,\sigma}^{\phantom\dagger} | B_3\rangle$ are momentum-dependent.
Momentum dependence can also arise from the $J$ term in $H$; for
example, the basis $B_2$ contains states with the hole translated by
$2a$ and a pair of flipped spins, and so matrix elements like $\langle
B_0|{\bf S}_i\cdot{\bf S}_j|B_2\rangle$ depend on {\bf k}.  The
results show odd-even effects in the order of the basis; as the basis
size increases, Trugman paths of higher order, and also states
differing from the starting state by nearest-neighbor spin
interchanges, are generated.

Related bases were also studied, in an effort to determine which
states are important for the hole properties.  The hopping basis can
be described symbolically as $B_n=\sum_{k=0}^n h^k |cN\rangle$ where
$h$ is the hole hopping operator.  We define also operators ${\cal
S}_8$, ${\cal S}_{12}$ and ${\cal S}_{20}$; the first scrambles the 8
spins at distances $a$ and $\sqrt2 a$ from the hole (giving 70 states
when operating on the $|cN\rangle$ state), the second these spins plus
the four at distance $2a$, and the third the 20 spins inside a
$5\times 5$ square minus the four corner sites.  If hole properties
like the bandwidth are determined primarily by configurations which
differ from the $|cN\rangle$ state only near the hole, then the bases
$\sum_k h^k{\cal S}_m|cN\rangle$, or (likely better) ${\cal
S}_m\sum_kh^k|cN\rangle$, should converge more rapidly than the
hopping basis; we find the opposite: when the bandwidth is plotted
against the inverse of the log of the basis size, these modified bases
behave like the hopping basis, except that properties are shifted
toward larger basis sizes.  We considered also two other bases, both
of which reduce the importance of string states (in which the hole
wanders without looping): (i) the basis $\sum_kM_m h^k|cN\rangle$
where the operator $M_m$ removes states in which the Manhattan
displacement ($|x|+|y|$) of the hole relative to its initial position
is greater than $ma$, and (ii) the basis $\sum_{k=0}^\infty (I_n h)^k
|cN\rangle$, where the operator $I_n$ removes states with more than
$n$ ``bad bonds'' (that is, it filters states according to their Ising
energy relative to the $|cN\rangle$ state; the limit $\infty$ means
that the hop-filter combination is applied until the basis no longer
grows, for given $n$).  Neither the Manhattan nor the Ising filters
improved the convergence.  We conclude from these numerical
experiments that the single-hole properties are determined not so much
by the spin configurations near the hole as by loop and string paths.
It appears that the hopping basis, whether in its original form or in
the modified forms we have investigated, is capable of only limited
accuracy even if carried to very high order.

%Implicit in the use of the hopping basis is the assumption that states
%omitted from the basis are unimportant for the dispersion, even though
%these states may be vital for the ground-state wavefunction and
%energy.  Examples are states differing from the $|cN\rangle$ state by
%a nearest-neighbor spin interchange far from the hole; these can be
%written as $S_i^+S_j^-|cN\rangle$ with $i$ and $j$ nearest-neighbor
%sites far from the hole. We have examined the amplitudes of such
%states which do appear in the basis (as large $|i|+|j|$ as permitted),
%finding that the amplitudes differ typically by less than a factor of
%2 between ${\bf k} = {\bf \Gamma}$ and ${\bf k} = {\bf S}$; this is
%comparable to the change in the amplitude of the $|cN\rangle$ state.
%These states do not appear to contribute greatly to the dispersion.

\section{Bandwidth}
\label{bandwidth}

Because the lattice is effectively infinite, the lowest energy can be
found for any $\bf k$. For both models, we found $E({\bf k})$ at 81
independent $\bf k$ values of the form $(2\pi n/L, 2\pi m/L)$ with $n$
and $m$ integers and $L = 32$, for $t/J$ values in the range $0.1\leq
t/J\leq10$ for the $t$-$J$ model and in the range $1\leq t/J\leq10$
for the strong-coupling model (for which the lower values of $t/J$ are
of little interest).

For the $t$-$J$ model, the energy is a minimum at ${\bf k}={\bf S}$
(and a maximum at ${\bf\Gamma}$) for $0.1\leq t/J\leq10$, for all
bases used ($B_6$ to $B_{13}$), in agreement with all previous work.

For the strong-coupling model, the energy is also a minimum at ${\bf
k}={\bf S}$ (and a maximum at ${\bf\Gamma}$) for all $t/J$ in the
range $1.0\leq t/J\leq10$, but only for the largest bases at small
$t/J$; this result disagrees with predictions (based on fits to
exact-diagonalization results for small
systems\cite{Fehske_1991_8473}) that the minimum is at ${\bf M}$ for
$t/J \leq 5$.  For the smaller bases, particularly for the smaller
values of $t/J$, the minimum can be at ${\bf M}$ or elsewhere along
the zone face; for example, the minimum is at ${\bf S}$ only in 11-th
order and higher for $t/J=1$.

\begin{figure}
\epsfxsize=3.5in \epsfbox{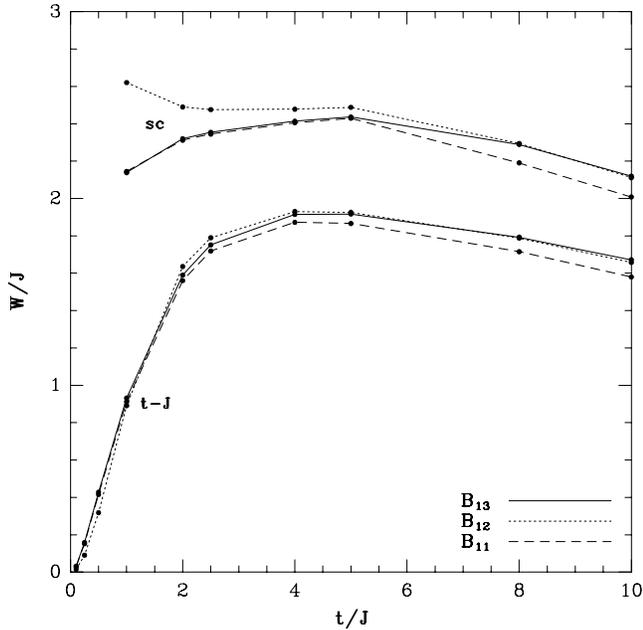}
\caption{Bandwidth $W$, in units of $J$, for the $t$-$J$ and
strong-coupling (sc) models as functions of $t/J$ for the three
largest bases used. The lines merely connect the points, here and in
following Figures.
\label{bw_ours_figure}
}
\end{figure}

Figure~\ref{bw_ours_figure} plots the bandwidth
$W=E({\bf\Gamma})-E({\bf S})$ for both models as found using the bases
$B_{11}$, $B_{12}$, and $B_{13}$.  The convergence is good for the
$t$-$J$ model at all $t/J$ investigated; it is moderately good for the
strong-coupling model at larger $t/J$, but worsens at smaller $t/J$.
The $t$-$J$ bandwidth is approximately $t$ for $t/J<2$ and $2J$ for
$t/J>2$, but decreases weakly at large $t/J$.  The strong-coupling
bandwidth is also about $2J$ (though about 20\% larger) and also
decreases as $t/J$ increases.  The hopping-basis results are
incompletely converged, however; the bandwidth is still increasing
with basis size, and the trend is greater at larger $t/J$.  It is
possible then that the slight decrease which we find is due to the
finite size of the hopping basis.

\begin{figure}
\epsfxsize=3.5in \epsfbox{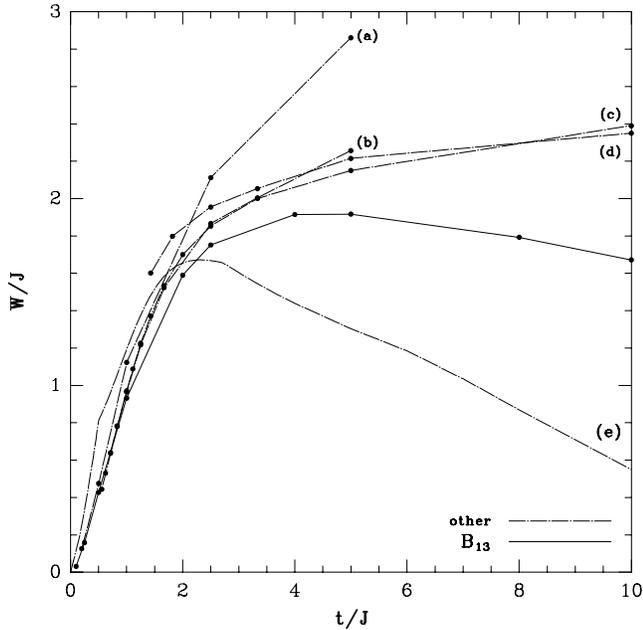}
\caption{Bandwidth $W$, in units of $J$, for the $t$-$J$ model as
functions of $t/J$.  The solid line gives the hopping-basis results
(for the basis $B_{13}$) and the dot-dash lines other results:
(a) Ref.\ \protect\onlinecite{Boninsegni_1992_4877},
(b) Ref.\ \protect\onlinecite{Marsiglio_1991_10882},
(c) Ref.\ \protect\onlinecite{Martinez_1991_317},
(d) Ref.\ \protect\onlinecite{Liu_1995_3156}, and
(e) Ref.\ \protect\onlinecite{Elrick_1993_6004}.
\label{bw_comp_figure}
}
\end{figure}

Figure~\ref{bw_comp_figure} compares our values for the $t$-$J$
bandwidth with those obtained by other methods; major differences
occur in the physical region $t/J>2$.  The hopping-basis results agree
best with loop-expansion
results\cite{Marsiglio_1991_10882,Martinez_1991_317,Liu_1995_3156}
and poorly with variational Monte-Carlo
results\cite{Boninsegni_1992_4877} (for unknown reasons); the
$4\times4$ exact-diagonalization results\cite{Elrick_1993_6004} at
large $t/J$ are unreliable due to finite-size effects.  Our results at
large $t/J$ are qualitatively consistent with the mean-field
result\cite{Schrieffer_1989_11663} $W\approx4J$ for strong coupling.

 From Figure 1, the normalized bandwidth difference
$(W_{sc}-W_{t\mbox{\small -}J})/J$ is almost independent of $t/J$ for
$t/J\agt4$.  Since $(H_{sc} - H_{t\mbox{\small -}J})/J=H_3/J$ has no
explicit dependence on $t$ or $J$, this suggests treating the
three-site terms as a perturbation to the $t$-$J$ model.
The error in the first-order result for the bandwidth difference
$\Delta W_1= \Delta E_1({\bf\Gamma}) - \Delta E_1({\bf M})$, where
$\Delta E_1({\bf k}) = \langle\Psi_{t-J}|H_3|\Psi_{t-J}\rangle({\bf k})$,
is less than 2\% at $t/J=10$ and $t/J = 8$, but is much larger at smaller
$t/J$ (52\% at $t/J=$ 4).
Of course the estimate for the strong-coupling bandwidth itself is much
better (errors are 0.3\%, 0.3\%, and 11\% at $t/J=$ 10, 8, and 4).
It appears then that the three-site terms can be treated in first order
for $t/J\agt6$.

Further investigation revealed that the first-order estimates of the
energy at $\bf S$ are excellent; $(\langle H_{sc}
\rangle_{t\mbox{\small -}J} - E_{sc})/W_{sc}$ is $0.1\%$, $0.09\%$,
$0.06\%$ and $0.04\%$ at $t/J=10$, 8, 4, and 1 respectively; the
corresponding values at $\bf\Gamma$ are $0.4\%$, $0.4\%$, $11\%$ and
$41\%$.  For unknown reasons, at intermediate $t/J$ values the
three-site terms appear to affect the $\bf\Gamma$ ground state
strongly and the ${\bf M}$ ground state very weakly.

\section{Dispersion}
\label{dispersion}

The Fourier coefficients $a_{lm}$ defined by
\begin{equation}
\label{dispersion_equation}
E({\bf k}) = \sum_{l,m=0}^{L/2} a_{lm} \cos{l k_x} \cos{m k_y}
\end{equation}
are easily obtained by inversion from the energy as a function of
${\bf k}$.  The symmetries of the lattice give $a_{lm} = a_{ml}$, and
$a_{lm} = 0$ for $l+m$ odd.  The independent coefficients are then the
81 $a_{lm}$ with $0\leq l\leq16$, $0\leq m\leq l$, and $l+m$ even.
The coefficient $a_{00}$ depends strongly on the order of the basis,
as more states important for the ground-state energy are generated; it
affects none of our results since we look only at quantities (like the
dispersion) which depend on energy differences.

\begin{figure}
\epsfxsize=3.5in \epsfbox{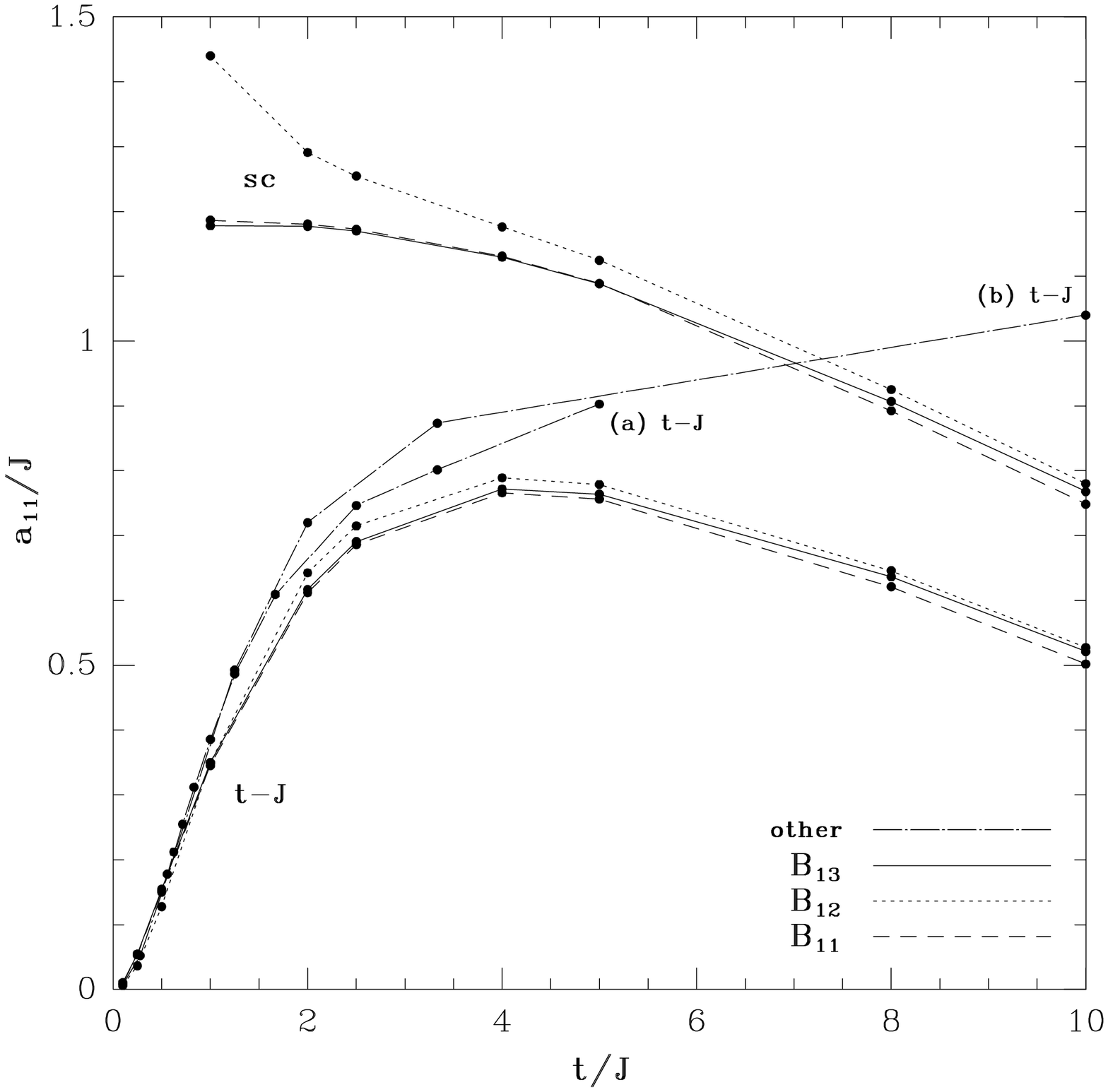}
\caption{The leading Fourier coefficient $a_{11}$, in units of $J$,
for the $t$-$J$ and strong-coupling (sc) models as functions of $t/J$
for the three largest bases used.  The dot-dash lines give other
results for the $t$-$J$ model:
(a) Ref.\ \protect\onlinecite{Marsiglio_1991_10882} and
(b) Ref.\ \protect\onlinecite{Martinez_1991_317}.
\label{a11_figure}
}
\end{figure}

\begin{figure}
\epsfxsize=3.5in \epsfbox{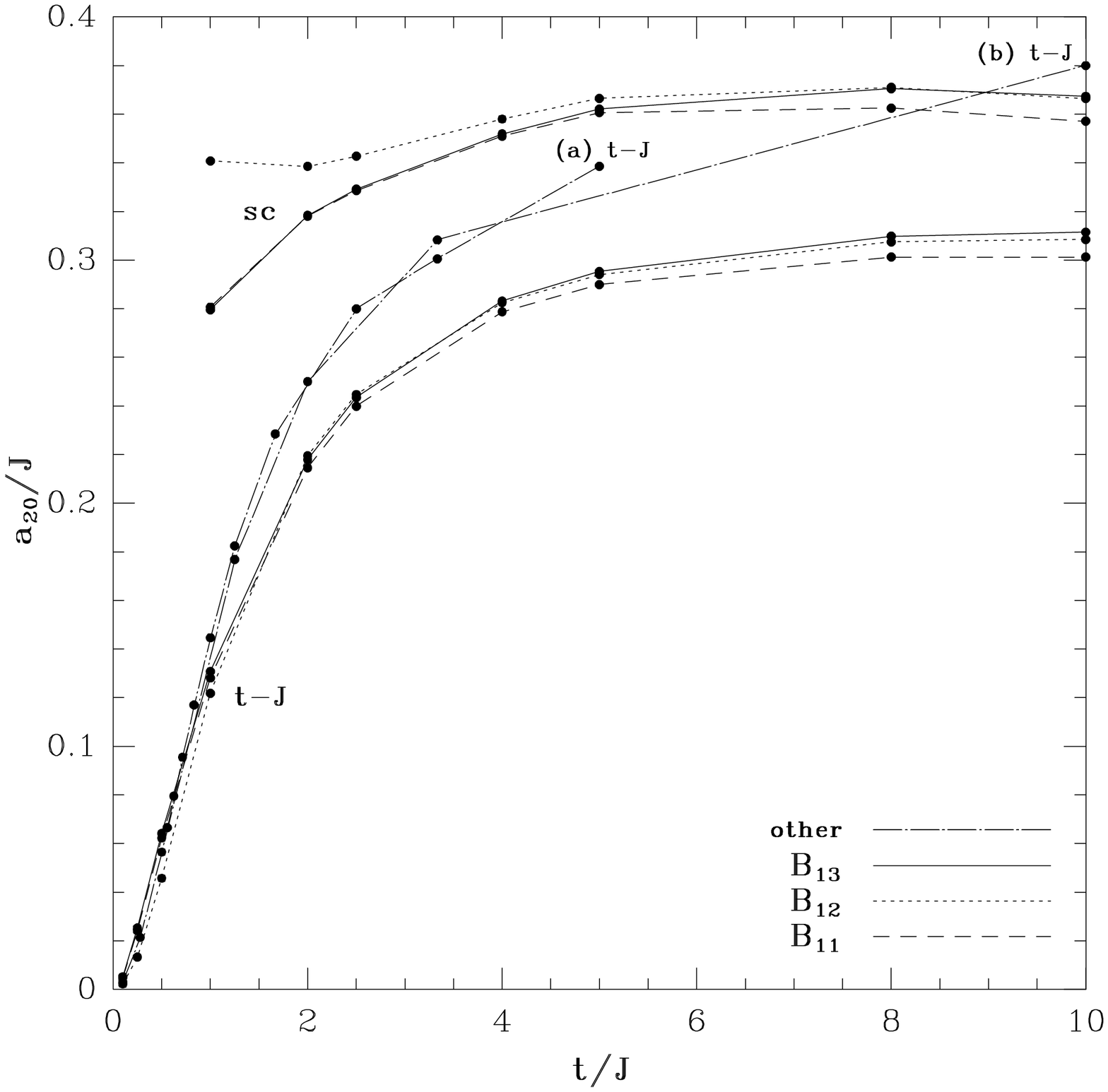}
\caption{The second leading Fourier coefficient $a_{20}$, in units of
$J$, for the $t$-$J$ and strong-coupling (sc) models as functions of
$t/J$ for the three largest bases used.  The dot-dash lines give other
results for the $t$-$J$ model:
(a) Ref.\ \protect\onlinecite{Marsiglio_1991_10882} and
(b) Ref.\ \protect\onlinecite{Martinez_1991_317}.
\label{a20_figure}
}
\end{figure}

Of the other coefficients, $a_{11}$ and $a_{20}$ (both positive) are
the largest, with the ratio $a_{20}/a_{11}$ less than about 0.6 for
both models for the range of $t/J$ values studied.  The remaining
coefficients are less than about $0.1a_{11}$ in magnitude for both
models at the $t/J$ values studied.  Figures \ref{a11_figure} and
\ref{a20_figure} plot the two leading coefficients as functions of
$t/J$ for the two models.  The convergence is of course qualitatively
the same as for the bandwidth, good for the $t$-$J$ model at all $t/J$
and for the strong-coupling model for $t/J\agt4$, but increasingly
poor for the latter with decreasing $t/J$.

At large $t/J$, the values $a_{20}/J$ are almost independent of $t/J$,
whereas the coefficients $a_{11}/J$ decrease with increasing $t/J$.
The strong-coupling coefficients are larger than the $t$-$J$
coefficients, reflecting the enhanced mobility due to the three-site
terms.  Also, at larger $t/J$, the difference
$(a_{20}/J)_{sc}-(a_{20}/J)_{t\mbox{\small -}J}$ for the two models is
almost independent of $t/J$, as is the difference in the values of
$a_{11}/J$, for the reason discussed in Section \ref{bandwidth}.
Figures \ref{a11_figure} and \ref{a20_figure} also plot other
results\cite{Marsiglio_1991_10882,Martinez_1991_317} for the $t$-$J$
Fourier coefficients; the agreement is as expected from Section
\ref{bandwidth}. Recent Monte-Carlo
results\cite{Dagotto_1994_728,Giamarchi_1993_2775}, available only at
$t/J = 2.5$, are about 25\% higher than ours.

\section{Band Masses}
\label{bandmass}

\begin{figure}
\epsfxsize=3.5in \epsfbox{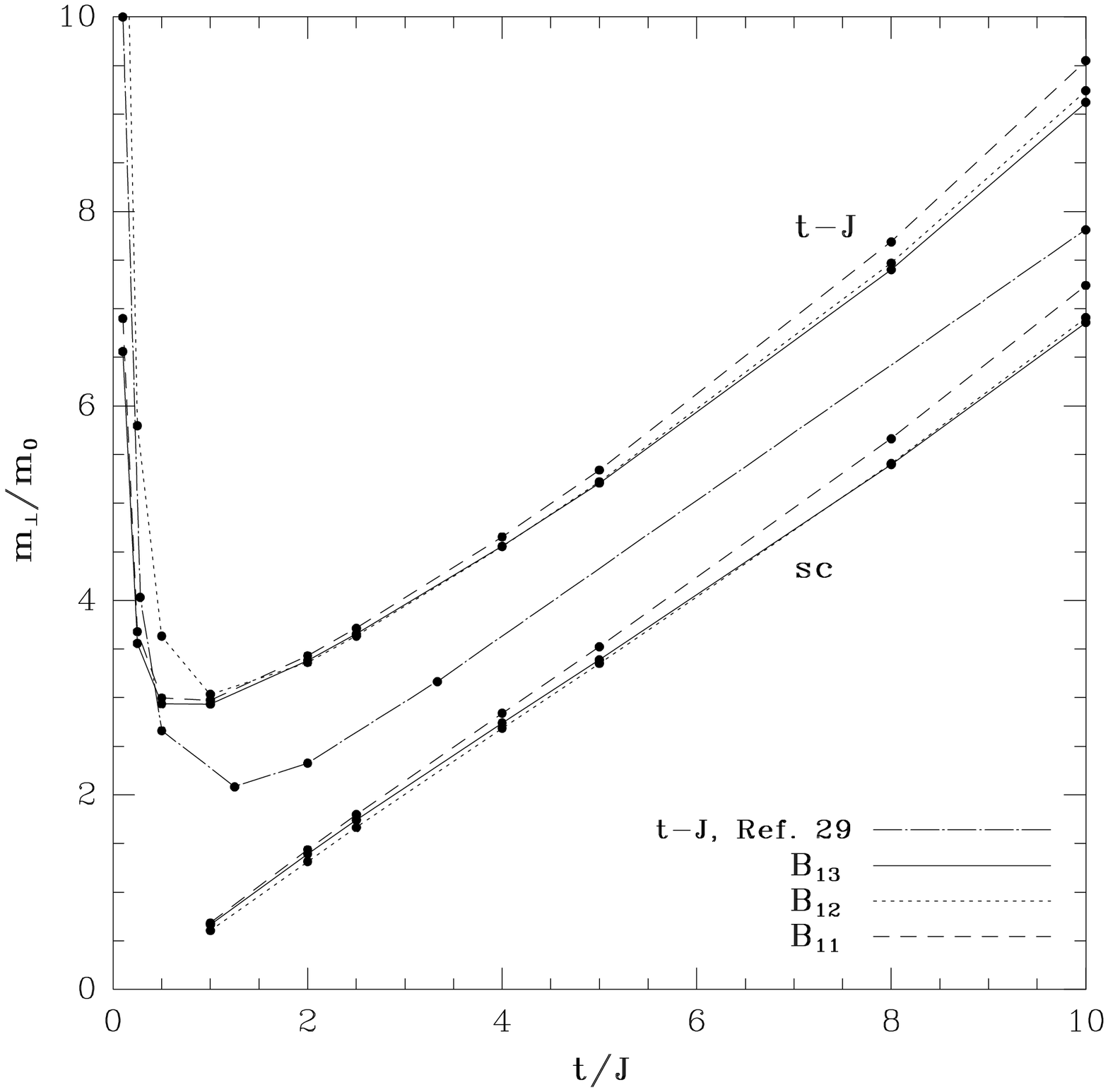}
\caption{Band mass perpendicular to the magnetic zone face at the band
minimum ${\bf k} = {\bf S}$, in units of the free band mass
$m_0=\hbar^2/{2t}$, for the $t$-$J$ and strong-coupling (sc) models as
functions of $t/J$.  The dot-dash line gives the $t$-$J$ results of
Ref.\ \protect\onlinecite{Martinez_1991_317}.
\label{mper_figure}
}
\end{figure}

\begin{figure}
\epsfxsize=3.5in \epsfbox{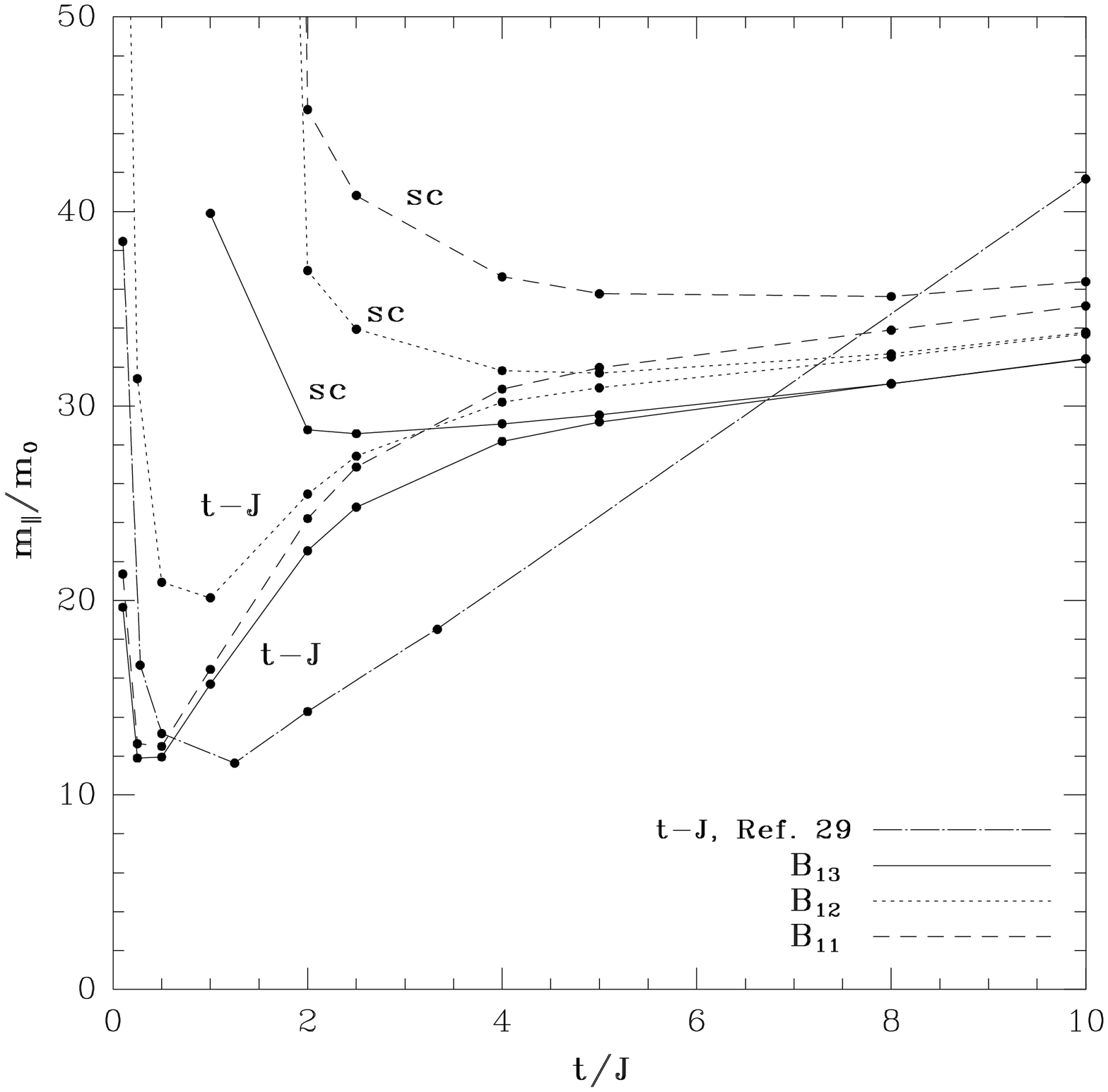}
\caption{Band mass parallel to the magnetic zone face at the band
minimum ${\bf k} = {\bf S}$, in units of the free band mass
$m_0=\hbar^2/{2t}$, for the $t$-$J$ and strong-coupling (sc) models as
functions of $t/J$.  The dot-dash line gives the $t$-$J$ results of
Ref.\ \protect\onlinecite{Martinez_1991_317}.
\label{mpar_figure}
}
\end{figure}

The band masses at the band minimum, which is at $\bf S$ for both
models in the region $1\leq t/J\leq10$, are defined in terms of the
second derivatives of $E({\bf k})$ with respect to ${\bf k}$:
\begin{equation}
 m_{\mu\nu} = \hbar^2 \left(\frac{\partial^2 E({\bf k})}
                        {\partial k_\mu\partial k_\nu}\right)^{-1}.
\end{equation}
The masses were obtained by calculating $E({\bf k})$ at additional
points near $\bf S$ and using finite-difference approximations for the
derivatives.  Figures~\ref{mper_figure} and \ref{mpar_figure} give
results for the masses perpendicular and parallel to the zone face
respectively, in units of the bare mass $m_0=\hbar^2/2t$.  The
parallel mass is much larger than the perpendicular mass, as found
previously\cite{Inoue_1990_2110,Macready_1991_5166,Fehske_1991_8473,%
Martinez_1991_317,Marsiglio_1991_10882,Dagotto_1994_728}.

The perpendicular mass is well converged for both models.  For the
$t$-$J$ model, $m_\bot/m_0$ is almost linear in $t/J$ at large $t/J$,
but flattens out at small $t/J$.  For the strong-coupling model,
$m_\bot/m_0$ is almost proportional to $t/J$; the smaller effective
mass reflects again the increased hole mobility relative to that in
the $t$-$J$ model.

The parallel mass is much more poorly converged, especially at smaller
$t/J$; even at $t/J=10$ (the most favorable value), the masses change
by over $5\%$ between the bases $B_{12}$ and $B_{13}$.  The poor
convergence results because the energies are nearly independent of
${\bf k}$ (the mass is large).  For large $t/J$, though, it appears
that $m_\parallel/m_0$ increases only weakly with $t/J$ for both
models and that the two models have the same parallel mass.

Figures~\ref{mper_figure} and \ref{mpar_figure} also give the results
from Ref. \onlinecite{Martinez_1991_317}, derived from their
dispersion results (Table II of Ref. \onlinecite{Martinez_1991_317})
using the free mass $m_0=\hbar^2/2t$, rather than the effective masses
of their Table III.  The difference
is due in part to a genuinely different dispersion, but part of it
arises because they used only two components in the Fourier expansion
(the parallel mass, being large, is particularly sensitive to small
changes in the energy).

\section{Spin-spin Correlations}
\label{correlations}

\begin{figure}
\epsfxsize=3.5in \epsfbox{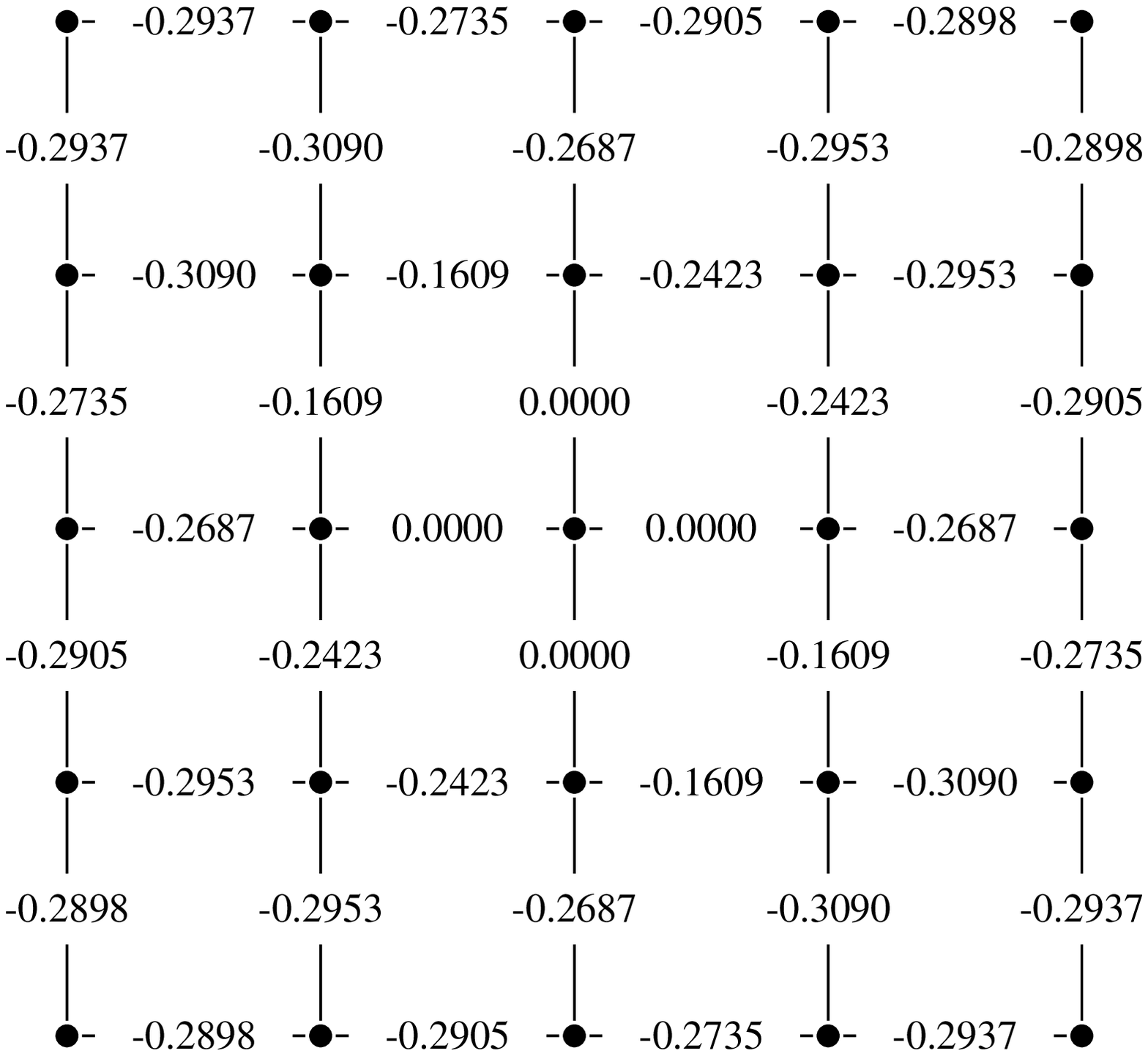}
\caption{Nearest-neighbor spin-spin correlations, in units of
$\hbar^2$, for the $t$-$J$ model at the band minimum ${\bf k} =
{\bf S}$ for $t/J = 2.5$ using the basis $B_{13}$.
\label{corr_tj_figure}
}
\end{figure}

\begin{figure}
\epsfxsize=3.5in \epsfbox{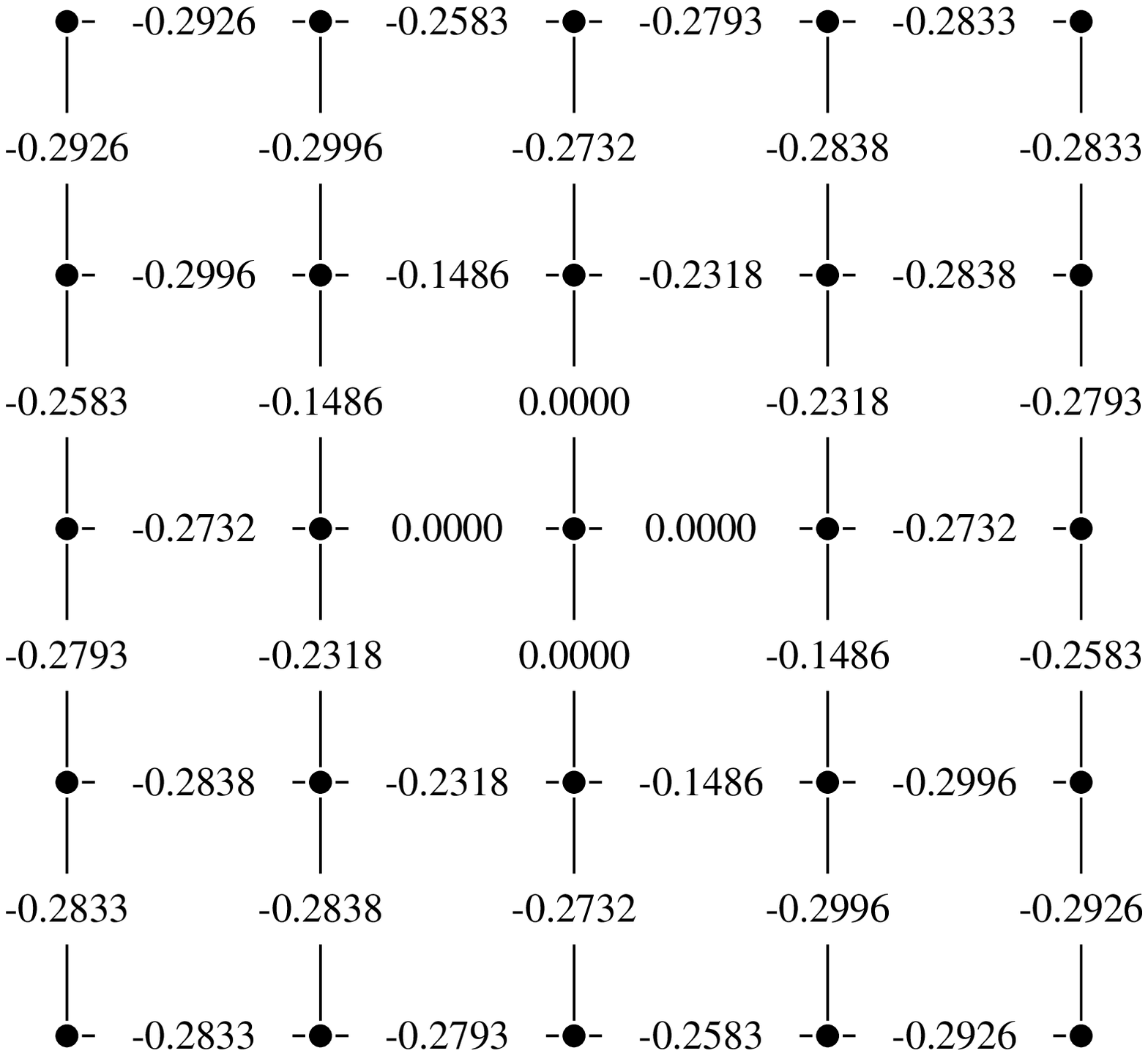}
\caption{Nearest-neighbor spin-spin correlations, in units of
$\hbar^2$, for the strong-coupling model at the band minimum ${\bf k}
= {\bf S}$ for $t/J = 2.5$ using the basis $B_{13}$.
\label{corr_sc_figure}
}
\end{figure}

Figures \ref{corr_tj_figure} and \ref{corr_sc_figure} show the
nearest-neighbor spin-spin correlation $\langle {\bf S}_i \cdot {\bf
S}_j \rangle$ for pairs of sites $i$ and $j$ near the hole, for the
$t$-$J$ model and strong-coupling model respectively.  The momentum is
${\bf k} = {\bf S}$ (the band minimum), $t/J = 2.5$, and the basis is
$B_{13}$.  In the units of $\hbar^2 = 1$ used, the spin-spin
correlation is -0.75 for a singlet pair of spins, $0.25$ for a triplet
pair, and -0.25 for a N\'eel pair.  The correlations are
antiferromagnetic, and moderately less than in the starting state.
The ``cigar'' polaron in Figures 7 and 8 is well known from other
studies\cite{Su_1988_9904,Dagotto_1989_6721,%
Elser_1990_6715,Inoue_1990_3467}.

\acknowledgments

This research was supported by the Natural Sciences and Engineering
Research Council of Canada.  Computations were done on a Kendall
Square Research 1-32 computer provided by University of Toronto
Instructional and Research Computing; we are grateful to UTIRC staff
for aid in making efficient use of the parallel architecture.

%%%%%%%%%%%%%%
% References %
%%%%%%%%%%%%%%

\eject
\narrowtext

%%%%%%%%%%%%%%%%%%%%%%
% Tables and Figures %
%%%%%%%%%%%%%%%%%%%%%%

\begin{table}[p]
\caption{Number of states in the hopping basis versus order of the basis.
\label{basissize_table}
}
\begin{tabular}{rrrr}
{\em Order} & {\em Number} &
{\em Order} & {\em Number} \\
{\em of Basis} & {\em of states} &
{\em of Basis} & {\em of states} \\ \hline
 0 &         1 &  8 &    9\,786   \\
 1 &         5 &  9 &   27\,990   \\
 2 &        17 & 10 &   80\,196   \\
 3 &        49 & 11 &  228\,196   \\
 4 &       141 & 12 &  650\,022   \\
 5 &       405 & 13 & 1\,842\,326 \\
 6 &    1\,177 & 14 & 5\,225\,938 \\
 7 &    3\,389 &    &             \\
\end{tabular}
\end{table}

\end{document}